\documentclass[showpacs,twocolumn,pra,superscriptaddress]{revtex4}
\usepackage{amsmath}
\usepackage{graphicx}
\usepackage{dcolumn}
\usepackage{color}

\usepackage{bm}
\newcommand{\tr}[1]{\,\mathrm{tr}\left\lbrace  #1 \right\rbrace}

\begin{document}
\title{Ion-trap simulation of the quantum phase transition in an exactly solvable model of spins coupled to bosons}

\author{Gian Luca Giorgi} \email{gianluca@ifisc.uib-csic.es}
\affiliation{Instituto de F\'{i}sica Interdisciplinar y Sistemas Complejos IFISC (CSIC-UIB),
Campus Universitat Illes Balears, E-07122 Palma de Mallorca, Spain}

\author{Simone Paganelli} \email{paganelli@ifae.es}
\affiliation{Grup de F\'{i}sica Te\`{o}rica, Universitat Aut\`{o}noma de Barcelona, E-08193 Bellaterra, Spain}

\author{Fernando Galve} 
\affiliation{Instituto de F\'{i}sica Interdisciplinar y Sistemas Complejos IFISC (CSIC-UIB),
Campus Universitat Illes Balears, E-07122 Palma de Mallorca, Spain}


\begin{abstract}
IIt is known that arrays of trapped ions can be used to efficiently 
simulate a variety of many-body quantum systems. 
Here, we show how it is possible to  build a model 
representing a spin chain interacting with bosons that is exactly solvable. 
The exact spectrum of the model at zero temperature and the ground-state 
properties are studied. We show that a quantum phase transition 
occurs when the coupling between spins and 
bosons reaches a critical value, which corresponds to a level crossing in the 
energy spectrum. Once the critical point is reached, the number of
 bosonic excitations in the ground state, which can be assumed as an order parameter,
 starts to be different from zero. The population of the bosonic mode is accompanied by
 a macroscopic magnetization of the spins. This double effect could represent 
a useful resource for the phase transition detection since a measure of the phonon
can give information about the phase  of the spin system.
A finite-temperature phase diagram is also given in the adiabatic
regime.
\end{abstract}
\maketitle

\section{Introduction}\label{intro}
Quantum spin chains play a fundamental role in the study of many-body systems
and quantum  phase transitions. These phenomena take place at 
zero temperature, that is in a purely quantum regime, and are induced by the variation of an internal parameter 
causing a critical change in the ground state due to level crossing in the energy spectrum \cite{sachdev}.
In recent years, a renewed interest in quantum phase transitions has been developed to understand the behavior
of entanglement near the critical point \cite{osborne,fazio,vidal,vedral}. 
On the contrary, 
the study of interactions between spins and light is considered as the starting point to introduce
 the framework of open systems, where the appearance of 
dissipation and decoherence can be understood \cite{weiss}. A prototypical example is the so-called spin-boson model
 \cite{leggett}, consisting of a single spin interacting with a multimode electromagnetic radiation,
 modeled as distribution  of quantum harmonic oscillators. Here, we discuss the case where a series of spins 
 (arranged in an isotropic $XY$ chain) interacts with one or more bosonic modes.

Despite the importance of spin systems, they are not always directly accessible 
experimentally, and efficient experimental quantum simulation methods are needed to study their properties.
It has been shown that spin chains can be efficiently simulated using 
internal and external degrees of freedom of trapped 
particles \cite{porras04,deng,ciaramicoli,johanning,lewenstein}. In particular, arrays of 
laser-cooled trapped ions seem to be very promising from an experimental point of view. 
Ions can be trapped with high spatial accuracy,
 and their internal states can be manipulated with high precision by means of 
the interaction with electromagnetic fields.  
The first experimental evidence of coupling between 
two-level systems, consisting in the transition from 
paramagnetic to ferromagnetic order in a two-spin quantum Ising model has recently been 
reported \cite{Friedenauer}. Furthermore, ion traps offer the possibility of engineering 
spin-boson coupling \cite{johanning,porras08}. 
Since spin chains are also important for practical applications, like quantum communication 
protocols \cite{bose}, their simulation would allow estimation of the decoherence 
effects which derive from the interaction with the environment \cite{noi}.

As said before, we propose a model which allows to observe the phase transition 
in an isotropic $XY$ spin chain coupled with external boson modes. The model we introduce in the following is exactly solvable. The phase transition manifests 
itself in a nonanalytical variation of the amplitude of the bosonic field.
In correspondence of this change, the  chain acquires a finite magnetization. The physical 
implementation of this model can be done within  the framework of trapped ions discussed before.

The paper is organized as follows. In Sec. \ref{II} we first introduce the general argument of simulation of spin chains with trapped ions.
 Then, we show how it is possible to simulate a particular model, consisting of a chain of spins in contact with a bosonic environment.
The critical properties at zero temperature  of this model are discussed in Sec. \ref{III},
while in Sec. \ref{IV}  we study the phase diagram at finite temperature 
in the case of a single slowly oscillating phonon. Finally, in Sec. \ref{V} we present our conclusions.

\section{Arrays of trapped ions and interaction between spins and bosons}\label{II}
As shown in Refs. \cite{porras04} and \cite{deng}, Coulomb chains in linear Paul traps can
simulate spin-spin interactions. An effective spin-spin Hamiltonian emerges as the result of the interaction of
the internal (electronic) states of ions with the 
phononic degrees of freedom generated by Coulomb repulsion. The total Hamiltonian includes a 
phonon bath ($H_\nu$), a state-dependent force produced by a set of lasers along the directions $\alpha=x,y,z$ ($H_f$), and an 
effective magnetic field that can be generated by forcing transitions between 
 the internal ion states ($H_m$). 
In units of $\hbar=1$, we have (see \cite{porras04})
\begin{eqnarray}
H_\nu&=&\sum_{\alpha,n}\omega_{\alpha,n}a^\dag_{\alpha,n}a_{\alpha,n},\nonumber\\
H_f&=&-2\sum_{\alpha,l}F_\alpha q_{\alpha,l}(1+\sigma^\alpha_l),\nonumber\\
H_m&=&\sum_{\alpha,l}B^\alpha\sigma^\alpha_l,
\end{eqnarray}
where $a_{\alpha,n}$ is the annihilation operator of the $n$ vibrational mode in the $\alpha$ 
direction of the phonon bath due to the Coulomb repulsion, $F_\alpha$ are the laser forces, $q_{\alpha,l}$ is the displacement, with respect to its equilibrium position, of ion $l$ in direction $\alpha$, and $B^\alpha$ is the effective 
magnetic field. Expressing the coordinates in terms of collective modes through the matrices $M^\alpha$,
\begin{equation}
 H_f=\sum_{\alpha,l,n}F_\alpha\frac{M^\alpha_{l,n}}{\sqrt{2m\omega_{\alpha,n}}}(a^\dag_{\alpha,n}+a_{\alpha,n})(1+\sigma^\alpha_l).
\end{equation}
A spin-spin Hamiltonian is obtained after the application of a suitable canonical 
transformation $U=e^S$, introduced to eliminate $H_f$: if 
\begin{equation}
 S=\sum_{\alpha,l,n}F_\alpha\frac{M^\alpha_{l,n}}{\sqrt{2m\omega^3_{\alpha,n}}}(a^\dag_{\alpha,n}-a_{\alpha,n})(1+\sigma^\alpha_l),
\end{equation}
the total system Hamiltonian turns out to be 
\begin{equation}\label{eqn:heff}
\tilde{H}=UHU^{\dag}=H_\nu+\frac{1}{2}\sum_{\alpha,l,j}J^\alpha_{l,j}\sigma_l^{\alpha}\sigma_j^{\alpha}+\sum_{\alpha,l}B^{\prime\alpha}\sigma_l^{\alpha}+H_E,
\end{equation}
where $B^{\prime\alpha}=B^{\alpha}-F_\alpha^2/(m\omega_\alpha^2)$, the coupling parameters $J$ are function of $F,M,\omega$, and $H_E$ 
is a residual term that 
can be neglected at low temperatures or by using highly anisotropic traps \cite{porras04}. 
A particular case of Eq. (\ref{eqn:heff}) is represented by the
isotropic $XY$ chain in the presence of a transverse field, numerically studied 
in Ref. \cite{deng}.

Here, we  show that a different interpretation of the unitary transformation allows us to obtain other kinds of phase 
transitions, involving, for instance, the interaction between spins and bosons. A possible way to do that 
consists in applying different transformations along the three axes. In fact, by limiting the 
transformation given before to the directions $x$ and $y$, through
\begin{equation}
 S_{x,y}=\sum_{\alpha=x,y}\sum_{l,n}F_\alpha\frac{M^\alpha_{l,n}}{\sqrt{2m\omega^3_{\alpha,n}}}(a^\dag_{\alpha,n}-a_{\alpha,n})(1+\sigma^\alpha_l),
\end{equation}
we obtain
\begin{eqnarray}
e^{-S_{x,y}}He^{S_{x,y}}&=&     
\sum_{l,n}F_z\frac{M^z_{l,n}}{\sqrt{2m\omega_{z,n}}}(a^\dag_{z,n}+a_{z,n})(1+\sigma^z_l) \nonumber\\
          &&+\frac{1}{2}\sum_{\alpha=x,y}\sum_{l,j}J^\alpha_{l,j}\sigma_l^{\alpha}\sigma_j^{\alpha}+\sum_{l}B^{z}\sigma_l^{z}\nonumber\\&&+
\sum_{\alpha=x,y}\sum_l B^{'\alpha}\sigma_l^{\alpha}+H_E^{\prime}+H_\nu,
\end{eqnarray}
Here, the correction $H_E^{\prime}$ can be neglected in the same limit of $H_E$.  In the case of a spatially
 homogeneous array, we can drop the dependence from the ion position in both $M$ and in $J$. 
Moreover, if the Coulomb interaction can be
considered as a perturbation with respect to the trapping potential (stiff limit), the trapping frequencies can be tuned to obtain relevant
spin-spin coupling only between nearest neighbors \cite{porras04}.
Along the transverse direction, a boson displacement can be applied to eliminate
the term $\sum_n F_z M^z_{n}/(\sqrt{2m\omega_{z,n}})(a^\dag_{z,n}+a_{z,n})$.
As a result,
\begin{eqnarray}
\tilde{H}&=&\sum_{\alpha=x,y}\frac{J^\alpha}{2}\sum_{l}\sigma_l^{\alpha}\sigma_{l+1}^{\alpha}  +\sum_{n}\frac{g_n^z}{\sqrt{N}}(a^\dag_{z,n}+a_{z,n})\sum_l\sigma^z_l\nonumber\\&&+ (B^{z}-2\sum_{n}\frac{g_n^{z2}}{\omega_n})\sum_l\sigma^z_l+\sum_{\alpha=x,y}\sum_l B^{'\alpha}\sigma_l^{\alpha}\nonumber\\&&+H_\nu-N\sum_{n}\frac{g_n^{z2}}{\omega_n},
\end{eqnarray}
where $g_n^z=F_z M^z_{n}\sqrt{(N/2m\omega_{z,n})}$. The last term is a constant that 
translates all the energies, and the fields $B^\prime$ can be set to zero. By fixing the external 
magnetic field $B^{z}=2\sum_{n}(g_n^{z2}/\omega_n)$, we arrive to
\begin{eqnarray}
\tilde{H}&=&\frac{1}{2}\sum_{l}(J^x\sigma_l^{x}\sigma_{l+1}^{x}+
J^y\sigma_l^{y}\sigma_{l+1}^{y}) \nonumber\\&&+\sum_{n}\frac{g_n^z}{\sqrt{N}}(a^\dag_{z,n}+a_{z,n})\sum_l\sigma^z_l+
H_\nu-\frac{NB^{z}}{2}\label{htilde}.
\end{eqnarray}
It is worth noting that, since the mutual distance between the ions makes any direct spin-spin interaction negligible, the spins we introduced are effective two-level systems that, because of the unitary
transformation,  are not trivially related to the original internal degrees of freedom of the ions. Nevertheless, the bosonic 
variables do refer to the real phonon modes along the $z$ direction, so their measurement appears to be the natural way
 to get information about the overall system.

\section{Critical properties}\label{III}
To describe the critical properties of this model, let us first discuss the simpler single-mode version. 
By assuming $J^x=J^y=J$, the Hamiltonian is
 \begin{equation}
 H_{sm}=J\sum_{l=1}^N[\sigma_{l}^{+}\sigma_{l+1}^{-}+H.c.]+\omega 
a^{\dag}a+\frac{g}{\sqrt{N}}(a^{\dag}+a)\sum_{l=1}^N\sigma_{l}^{z},\label{acca}
\end{equation}
where $N$ is the total number of spin, and where periodic boundary conditions are imposed.
As it is useful to work with adimensional quantities, we use $J$ as energy unit and define the rescaled phonon
frequency $\gamma=\omega / J$ as well as the bare coupling constant $\lambda=g^2/(J \omega)$.  Using this notation
the Hamiltonian reads 
\begin{equation}
H_{sm}=\sum_{l=1}^N[\sigma_{l}^{+}\sigma_{l+1}^{-}+H.c.]+\gamma 
a^{\dag}a+\sqrt{ \frac{\lambda \gamma }{N}}(a^{\dag}+a)\sum_{l=1}^N\sigma_{l}^{z}.\label{acca_ad}
\end{equation}
To diagonalize it, first we introduce the
Jordan-Wigner transformation \cite{lieb}, 
defined by $\sigma _{l}^{z}=1-2c_{l}^{\dagger }c_{l}$, $\sigma
_{l}^{+}=\prod_{j<l}\left( 1-2c_{j}^{\dagger }c_{j}\right) c_{l}$, and $\sigma _{l}^{-}=\prod_{j<l}\left( 1-2c_{j}^{\dagger }c_{j}\right)
c_{l}^{\dagger }$, mapping spins into spinless fermions. Then, the Fourier transform
\begin{equation}
c_k=\frac{1}{\sqrt{N}}\sum_{l=1}^N e^{i2\pi kl/N}c_l
\end{equation}
allows us to write, in the thermodynamic limit,
\begin{equation}
 H_{sm}=\sum_{k}\epsilon_k c_k^\dag c_k+\gamma a^{\dag}a+\sqrt{ \frac{\lambda \gamma}{ N}}(a^{\dag}+a)\sum_{k}( 1-2c_{k}^{\dagger }c_{k}),
\end{equation}
with $\epsilon_k=-2\cos (2\pi k/N)$. Fermions can be decoupled from the bosonic degree of freedom by 
applying the displacement operator $D=\exp{[\hat{\alpha}(a^\dag-a)]}$ such that $DaD^{\dag}=b$, 
$\hat{\alpha}=-\sqrt{\lambda/(\gamma  N)}\sum_{k}( 1-2c_{k}^{\dagger }c_{k})$ being an operator in the fermionic space. 
Since $[\hat{\alpha},\sum_{k}\epsilon_k c_k^\dag c_k]=0$, fermions are not transformed into polarons because of 
the application of $D$. Moreover, it is immediately seen that $[b,b^\dag]=1$. This displacement replaces the boson vacuum with a coherent state
whose amplitude is determined by the transverse component of the total spin.
The final Hamiltonian is
\begin{equation}
\tilde{H}_{sm}=\sum_{k}\epsilon_k c_k^\dag c_k+\gamma b^{\dag}b-\frac{\lambda}{N}[\sum_{k}(1-2c_{k}^{\dag}c_{k})]^{2}. \label{hd}
\end{equation}
Adding the boson displacement $D$ to $S_{z}$ and $S_{x,y}$ of Sec. \ref{II} amounts to building the total $S$ of Ref. \cite{porras04}
for a particular choice of parameters. If one disregards the term $\gamma b^{\dag}b$, the Hamiltonian \ref{hd}
 can now be interpreted as a $XXZ$ spin chain with short-range coupling along the radial direction
and axial long-range coupling. These kinds of interaction ranges are realistic, since radial modes can mediate nearest-neighbor spin-spin interactions
(stiff limit), and transverse modes are known to generate
long-range interactions (soft limit). While the range of interactions along the transverse direction
depends only on the number of ions and cannot be manipulated by means of laser fields, the stiff 
limit in the radial direction can be reached by increasing the trapping frequencies \cite{deng}.
Performing the unitary transformations in two separate steps helped us to interpret this model 
in terms of the coupling with a real phonon. Without performing the two steps separately, we could not discuss the spin-boson phase transition.


At first sight, $\tilde{H}_{sm}$ is similar to that of the $XX$ chain in a transverse field \cite{katsura}, whose 
Hamiltonian in the Jordan-Wigner space would be $\sum_{k}\epsilon_k c_k^\dag c_k-h\sum_{k}(1-2c_{k}^{\dag}c_{k})$. While the latter 
model has  the $U(1)$ symmetry, since it is invariant under rotations around the $z$ axis, the Hamiltonian (\ref{acca}) has an 
additional symmetry, also being invariant under the action of the operator ${\cal S}=\prod_l\sigma_{l}^{x}\otimes\exp{[i\pi a^{\dag}a]}$.
The $U(1)$ symmetry implies, in the language of fermions, that eigenstates of $H$ have a fixed number of particles.
The symmetry ${\cal S}$ could be broken, for instance, by adding a displacing bosonic field, $a+a^\dag$, and then sending it to $0$. 
 The average 
value of $a+a^\dag$ can be assumed as an order parameter to study the breakdown of ${\cal S}$.

Since we are interested in the implementation in an array of ion traps, finite-size effects should be taken into account. 
A detailed analysis of $XYZ$ models in finite cycles has been made in several papers \cite{katsura,prb,depa}. In our model, 
after the Jordan-Wigner transformation, the Hamiltonian becomes
\begin{eqnarray}
 H_{sm}&=&\sum_{l=1}^{N-1}(c_{l+1}c_{l}^{\dag}+c_{l}c_{l+1}^{\dag})-
{\cal P}(c_{1}^{\dag}c_{N}+c_{N}^{\dag}c_{1})+\gamma a^{\dag}a\nonumber\\&&+\sqrt{ \frac{\lambda \gamma}{ N}}(a^{\dag}+a)\sum_{k}( 1-2c_{k}^{\dagger }c_{k}),
\end{eqnarray}
where $N$ is assumed to be an even number and where ${\cal P}=\prod_{l=1}^{N}(1-2c_{l}^{\dag}c_{l})$. Its possible 
eigenvalues are $\pm1$. Note that ${\cal P}$ is a measure of the parity of the number of particles of each state. 
Since $[H_{sm},{\cal P}]=0$, all eigenstates of $H_{sm}$ have definite parity, and we can proceed to
a separate diagonalization of $H_{sm}$ in the two subspaces corresponding
to ${\cal P}=\pm 1$. Then, if we introduce the two new Hamiltonians $H_{sm}^{\pm}=H_{sm}({\cal P}=\pm1)$, the complete set of 
eigenvectors of $H_{sm}$
will be given by the odd eigenstates of $H_{sm}^{-}$ and the even eigenstates of $H_{sm}^{+}$. The choice of ${\cal P}=-1$ or ${\cal P}=+1$ 
amounts to having, respectively, periodic or antiperiodic boundary conditions. This implies that $H_{sm}^{+}$ and $H_{sm}^{-}$ are diagonalized 
through two Fourier transforms which differ from each other because of the set of allowed values of $k$. 
If half-integer values ($k=1/2,3/2,\ldots, N
-1/2$) are used to diagonalize $H_{sm}^{+}$, in the case of $H_{sm}^{-}$ we have $k=1,2,\ldots, N$. Now, the displacement operator $D$ can 
be introduced, and a structure formally identical to that given in Eq. (\ref{hd}) can be done for both $H_{sm}^{+}$ and $H_{sm}^{-}$, with the 
proper choice of $k$.

The ground state of the bosonic part can be identified as the vacuum state of the operator $b$, and it corresponds, in the 
original representation, to a coherent state of amplitude equal to the average value of the operator $\hat{\alpha}$ calculated 
on the fermionic ground state, whose structure is now enlightened. Since both $H_{sm}^{+}$ and $H_{sm}^{-}$ are exactly solvable, their 
spectra can be calculated by measuring the energy of all possible configurations, obtained adding the desired number of fermions. 
The energy of a state with $m$ particles is 
\begin{equation}
 E_{{\cal M}}=-\frac{\lambda(N-2m)^2}{N}-2\sum_{k\in {\cal M}}\cos \frac{2 \pi k}{N},\label{mm}
\end{equation}
where ${\cal M}$ is a string representing the occupation of different modes. The sum is extended only to the values of $k$ where a 
particle is present.
For large values of $\lambda$, the first term will be larger than the sum, unless $N-2m$ is very small. The ground state 
corresponds to the sequences which maximize $(N-2m)^2$, that is, to $m=0$ or $m=N$, and it is twofold degenerate. In fact, both the 
state $|\Phi^{+}\rangle=|0\rangle\otimes|\alpha\rangle$ and the state $|\Phi^{-}\rangle=\prod_{k}c_{k}^{\dag}|0\rangle\otimes|-\alpha\rangle$ 
have energy equal to $E_0=-N\lambda$. Here the state $|\Phi^{+}\rangle$ is, in the original fermion-boson representation (before the application of $D$), the tensor 
product of the fermionic vacuum and of a coherent bosonic state of amplitude $\alpha=-g\sqrt{N}/\omega$, while in $|\Phi^{-}\rangle$ 
the fermionic system is fully occupied, and the boson coherent state has amplitude $-\alpha$. 
As $\lambda$ decreases, the two terms in Eq. (\ref{mm}) start to compete with each other. 
Let us call $\lambda_m$ the value such that $E_0=\min\{E_{{\cal M}}\}$. Since, by symmetry, $\lambda_m=\lambda_{N-m}$, 
we limit out consideration $m=1,2,\dots,N/2$. It can be shown that, if $m<m'$, then $\lambda_m<\lambda_{m'}$. 
In other words, the first ground-state level crossing takes place between $|\Phi^{+}\rangle$ (or $|\Phi^{-}\rangle$) 
and a state with $N/2$ particles. This transition occurs at the value of $\lambda_{N/2}=\lambda_c$ determined by
\begin{equation}
N \lambda_c=4 \sum_{k=1}^{N/4}\cos\frac{(2k-1)\pi}{N}.
\end{equation}
In the thermodynamic limit, performed replacing the sum with an integral, we obtain $\lambda_c=2/\pi$. 
As a consequence of the half-filling, $\sum_{k}( 1-2c_{k}^{\dagger }c_{k})=0$, and the boson part is left in its vacuum. 
This state can be written as $|\Phi^{HF}\rangle=\prod_{|k|<N/4}c_{k}^{\dag}|0\rangle\otimes|0\rangle$.

\begin{figure}
    \includegraphics[height=5cm]{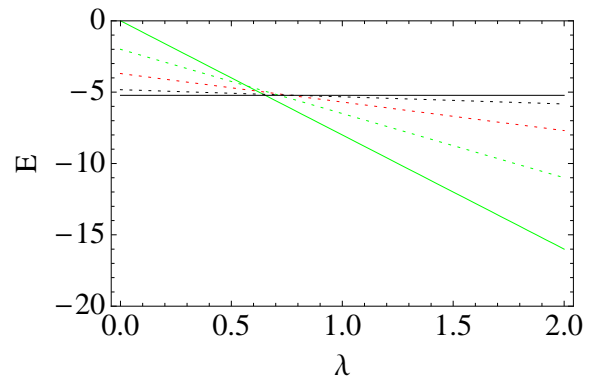}\\
  \caption{Ground-state energy of $H$ as a function of $\lambda$ for a chain of 8 spins. 
Both the energy and $\lambda$ are measured in units of $J$ or $\omega$ (we assume $J=\omega=1$). 
For any given number of particles, only the lower level is plotted. The solid green line represents the energy of $|\Phi^{+}\rangle$ 
and $|\Phi^{-}\rangle$, while the solid black line corresponds to $|\Phi^{HF}\rangle$. As we can see, there is only one transition, 
occurring at $\lambda_c\simeq 0.65$. All energies derived from numbers of fermions different from $0,N/2,N$ are excitations. 
Specifically, the dotted green line represents the energy of the one-particle state, the dotted red line is for the two-particle state, and the dotted black 
line corresponds to the three-particle state. Due to the particle-hole symmetry, states with $m$ and $N-m$ particles have 
the same energy.}
\label{energie}
\end{figure}

By further decreasing $\lambda$, no other transitions are observed. 
Then the state $|\Phi^{HF}\rangle$ is the Hamiltonian ground state for every $\lambda<\lambda_c$. 
It is worth noting that both $|\Phi^{HF}\rangle$ and $ |\Phi^{+}\rangle$ (or $ |\Phi^{-}\rangle$) are eigenstates of $H_{sm}^+$. 
In any case, eigenstates of $H_{sm}^-$ are excitations. In Fig. \ref{energie}, we plot the lowest energy levels for a chain of eight spins. 
The transition $ |\Phi^{\pm}\rangle\rightarrow |\Phi^{HF}\rangle$ is observed.

While $|\Phi^{HF}\rangle$ is an eigenstate of the symmetry operator ${\cal S}$, this is not true in the case of 
$|\Phi^{+}\rangle$ and $|\Phi^{-}\rangle$. In fact, ${\cal S}|\Phi^{\pm}\rangle=|\Phi^{\mp}\rangle$. 
The possibility of obtaining a coherent emission of light on the bosonic mode is inherently linked to 
the breakdown of the Hamiltonian symmetry, since the only coherent state which is also an
eigenstate of $\exp{[i\pi a^{\dag}a]}$ is the vacuum. While symmetry breaking takes place 
independently of the system size (the number of spins), only in the thermodynamic limit does a true phase 
transition take place, since the Hilbert spaces that can be built up starting from $|\Phi^{+}\rangle$ 
and $|\Phi^{-}\rangle$ become unitarily nonequivalent.

The exact results we propose have been obtained in the presence of periodic boundary conditions.
By releasing this hypothesis, they are correct only in the thermodynamic limit, 
while in the case of finite size, as in the example in Fig. \ref{energie}, corrections to the energy levels are expected.
Actually, the structure of eigenfrequencies giving rise to the phase transition from $|\Phi^{\pm}\rangle$ to $|\Phi^{HF}\rangle$ can be observed also
in the case of a very short open chain ($N=4$).

In the presence of multimode radiation,
Hamiltonian (\ref{htilde}) can be recovered modifying Hamiltonian (\ref{acca}) in the following way
\begin{eqnarray}
 H_{mm}&=& \sum_{l=1}^N[\sigma_{l}^{+}\sigma_{l+1}^{-}+H.c.]+
\sum_n\gamma_n a^{\dag}_n a_n\nonumber\\&&+\sum_n\sqrt{\frac{\lambda_n \gamma_n}{N}}(a_n^{\dag}+a_n)\sum_{l=1}^N\sigma_{l}^{z},\label{accamm}
\end{eqnarray}
and it is subject to the symmetry ${\cal S}_{mm}=\prod_l\sigma_{l}^{x}\otimes\exp{[i\pi\sum_n a_n^{\dag}a_n]}$. 
The diagonalization is performed through the application of the Jordan-Wigner transformation and of the operator 
$\prod_n D_n$, with $D_n=\exp{[\hat{\alpha_n}(a_n^\dag-a_n)]}$, and 
$\hat{\alpha_n}=-\sqrt{\lambda_n/(\gamma_n N)}\sum_{k}( 1-2c_{k}^{\dagger }c_{k})$. 
As a result, we have
\begin{equation}
 \tilde{H}_{mm}=\sum_{k}\epsilon_k c_k^\dag c_k+
\sum_n\omega_n b^{\dag}_n b_n-\frac{\Lambda}{N}[\sum_{k}(1-2c_{k}^{\dag}c_{k})]^{2},\label{hdm}
\end{equation}
where $\Lambda=\sum_ng_n^2/(J \omega_n)$ is related to the spectral density of the bath. 
In Ref. \cite{porras08}, the authors showed how baths characterized by different spectral densities can be simulated. 
Being $\tilde{H}_{mm}$ formally identical to $\tilde{H}_{sm}$, we expect a phase transition for $\Lambda_c=2/\pi$. For $\Lambda>\Lambda_c$, 
the two degenerate ground states are $|\Phi^{+}_{mm}\rangle=|0\rangle\otimes\prod_n|\alpha_n\rangle$ and 
$|\Phi^{-}_{mm}\rangle=\prod_{k}c_{k}^{\dag}|0\rangle\otimes\prod_{n}|-\alpha_n\rangle$, while for  
$\Lambda<\Lambda_c$ the non degenerate half filled ground state 
is $|\Phi^{HF}_{mm}\rangle=\prod_{|k|<N/4}c_{k}^{\dag}|0\rangle\otimes\prod_{n}|0\rangle$.

To clarify the proposal, we comment on the preparation scheme needed to observe phase transitions in this system. As mentioned in \cite{porras04}, the procedure consists
 in initialization to the state $\left|\downarrow...\downarrow\right\rangle$, and then adiabatic switching of the spin-spin couplings. This process is repeated
several times for different ratios of $J/g$, while the phase transition is detected through fluorescence of the ions internal levels, as experimentally realized in \cite{Friedenauer}.
Here, we also need to perform motional ground-state cooling as far as axial degrees of freedom are concerned.
 The possible errors in simulating this model with trapped ions comes from the canonical transformation and $H_E'$, where the error is quantified by the boson displacements
 and the mean phonon number (given by temperature) of the neglected transverse modes. Therefore they have to be cooled down too \cite{porras04}. 

\section{Temperature effects in the adiabatic limit}\label{IV}
In this section we analyze the phase transition at finite  temperature  in the so-called adiabatic 
limit ($\gamma\ll 1$) with a single coupled mode.
This limit corresponds to the case of a very slowly oscillating phonon and it is asymptotically 
exact at high temperatures.
The adiabatic  Hamiltonian is obtained by neglecting the kinetic energy of the phonon and treating
the coordinate as a parameter  
\begin{equation}
H_{AD}=\sum_{l=1}^N [\sigma_{l}^{+}\sigma_{l+1}^{-}+H.c.]+\frac{\nu^2}{4 \lambda}+
\frac{\nu}{\sqrt{N}}\sum_{l=1}^N\sigma_{l}^{z},
\end{equation}
where we introduced the adimensional coordinate $\nu=(g/J)\sqrt{2 \omega m} x$.
The Hamiltonian is a isotropic $XY$ model in an external field, parametrically dependent 
on $\nu$, it can be diagonalized in the same 
way as described previously  and  reduced in two blocks corresponding to 
odd and even pseudofermion occupation number 
\begin{equation}
H_\pm=\sum_k \epsilon^{AD}_k(\nu) c^\dag_k c_k +\frac{\nu^2}{4 \lambda}+\frac{\nu}{\sqrt{N}},
\end{equation}
with $\epsilon^{AD}_k(\nu)=-2\left[\cos(2 \pi k/N)+\nu/\sqrt{N}\right]$.
The energy of a configuration with a filling $m$  is
\begin{equation}
E_{{\cal M}}=-2\left(\frac{\nu m}{\sqrt{N}}+\sum_{k\in{\cal M}}\cos\frac{2 \pi}{N}k\right).
\end{equation}
The partition function, defined by
\begin{equation}
 Z\propto \int d\nu Z_0(\nu,\beta) \exp{- \beta (\frac{\nu^2}{4 \lambda}+\frac{\nu}{\sqrt{N}})},
\end{equation}
with the adimensional parameter $\beta=J/k_B T$ and where
\begin{equation}
Z_0(\nu,\beta)=\tr{e^{-\beta \sum_k \epsilon^{AD}_k(\nu) c^\dag_k c_k}},
\end{equation}
can be written as   $Z \propto \int d\nu \exp{- \beta V_{AD}(\nu,\beta)} $
so defining  an adiabatic potential
\begin{equation}
V_{AD}(\nu,\beta)=\frac{\nu^2}{4 \lambda}+\frac{\nu}{\sqrt{N}}+
-\frac{1}{\beta} \ln Z_0(\nu,\beta).
\end{equation}

\begin{figure}
\includegraphics[height=6cm,angle=-90]{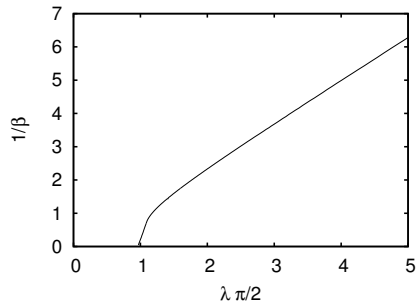}\\
\caption{Phase diagram in the adiabatic approximation. Data were obtained with 10 spins and
numerical evaluation of the adiabatic potential. 
The left region corresponds to a single-well adiabatic potential, 
the right region to a double-well potential.}
\label{fig:ph_diag}
\end{figure}

Above the critical temperature, $V_{AD}(\nu,\beta)$ is minimized by $\nu=0$, while, 
at the critical point, 
the potential admits two minima, allowing for a displacement of the phonon. The phase diagram 
obtained is plotted in Fig. \ref{fig:ph_diag}. 

The experimental procedure to measure the phase transition is exactly as in the zero-temperature case. 
The radial vibrations should be cooled, while the axial degrees of freedom can be prepared in a ``thermal state''.
The temperature of this state can be controlled by reducing the power of the laser 
(resolved sideband)  cooling mechanism. To raise it further, the power of the
Doppler laser cooling has to be tuned. 
We still expect to observe an abrupt change in fluorescence properties, as in the zero-temperature case, since for moderate temperatures
the coherent state $|\alpha\rangle$ becomes thermal, but the mean number of phonons does not disappear abruptly. 
According to Fig. \ref{fig:ph_diag}, the order parameter value should change as a function of $T$. 

\section{Conclusions}\label{V}
To resume, we have shown that a system of trapped ions can be mapped into an isotropic $XY$ chain interacting with phonons, following 
the scheme in Ref. \cite{porras04}.  Modifying the original canonical transformations needed to write a 
spin-spin Hamiltonian,  the vibrational
degrees of freedom along a fixed direction are coupled with the effective spins the other degrees of freedom are mapped onto. 
The resulting model is exactly solvable and exhibits a quantum phase transition, due to the breaking of a symmetry which takes 
into account both the spin and the boson degrees of freedom. In this phase, phonons are in a coherent state with
finite amplitude. Since the phonons  refer to the real ionic vibration  and not to 
an effective quantity, phase transition detection should be possible using state-of-the-art techniques.
The experimental realization of the building block of such 
simulations (Ref. \cite{Friedenauer}), encourages this kind of investigation.

\acknowledgments
GLG acknowledges discussion with F. de Pasquale.
SP acknowledges discussion with A. Sanpera and G. de Chiara.
This work was partially funded by CoQuSys project (200450E566). GLG and SP are supported by the Spanish Ministry of Science and Innovation 
through the program Juan de la Cierva.

\end{document}